\documentclass[aps,prl,twocolumn,groupedaddress,psfig]{revtex4}
\tighten

\usepackage{epsfig}

\newcommand{\fr}[2]{\frac{#1}{#2}}

\newcommand{\la}{\lambda}

\newcommand{\be}{\begin{equation}}
\newcommand{\ee}{\end{equation}}
\newcommand{\bea}{\begin{eqnarray}}
\newcommand{\eea}{\end{eqnarray}}

\def\ga{\mathrel{\raise.3ex\hbox{$>$\kern-.75em\lower1ex\hbox{$\sim$}}}}
\def\la{\mathrel{\raise.3ex\hbox{$<$\kern-.75em\lower1ex\hbox{$\sim$}}}}

\newcommand{\eg}{{\it e.g.,}\ }

\newcommand{\coupling}{\kappa} 

\newcommand{\bbounda}{430}
\newcommand{\bboundb}{510}
\newcommand{\bboundc}{1.1}
\newcommand{\bboundd}{1.5}

\newcommand{\babar}{\mbox{\sc{BaBar}}}

\def\sss{\scriptscriptstyle}
\def\ms{m_{\sss S}}

\def\EW{{\sss EW}}

\def\mht{m_{\tilde{h}}}

\def\vew{v_\EW}
\def\mmi{m_{\rm min}}

\begin{document}

\title{Search for Dark Matter in $b\rightarrow s$ Transitions with 
Missing Energy}
\author{Chris Bird$^{1}$, Paul Jackson$^{1}$, Robert Kowalewski$^{1}$,
 and Maxim Pospelov$^{1,2}$}
\affiliation{$^{1}$ Department of Physics and Astronomy,
University of Victoria, Victoria, BC, V8P 1A1, Canada}
\affiliation{$^{2}$ Centre for Theoretical Physics, University of Sussex, 
Brighton
BN1 9QJ,~~UK}

\begin{abstract}
Dedicated underground experiments searching for dark matter have little 
sensitivity to GeV and sub-GeV masses of dark matter particles. 
We show that the decay of $B$ mesons to $K(K^*)$ and missing energy in 
the final state can be an efficient probe of dark matter models in this mass range. 
We analyze the minimal scalar dark matter model to show that the 
width of the decay mode with two dark matter scalars 
$B\rightarrow KSS$ may exceed the decay width in 
the Standard Model channel, $B\rightarrow K\nu\bar\nu$, by up to two orders 
of magnitude. Existing data from $B$ physics experiments almost entirely exclude dark matter
scalars with masses less than 1 GeV.  Expected data from B factories will probe the range of 
dark matter masses up to 2 GeV.  
\end{abstract}

\maketitle

\section{Introduction}

Although the existence of dark matter is firmly established through its 
gravitational interaction, the identity of dark matter remains a big mystery. 
Of special interest for particle physics are models of 
weakly interacting massive particles (WIMPs), which have a number of 
attractive features: well-understood mechanisms of ensuring the correct 
abundance through the annihilation at the freeze-out, milli-weak to 
weak strength of couplings to the ``visible'' sector of the Standard Model (SM), 
and as a consequence, distinct possibilities for WIMP detection. 
The main parameter governing the abundance today is WIMP annihilation 
cross section directly related to the dark matter abundance.
In order to keep WIMP abundance equal or smaller than the observed dark 
matter energy density, the annihilation cross section has to satisfy the lower
bound, $\sigma_{\rm ann} v_{\rm rel} \ga 1$ pb, (see {\em e.g.} \cite{kolb-turner}). 
In all WIMP models studied to date the annihilation cross section 
is suppressed in the limit of very large or very small mass of a 
WIMP particle $S$.  This confines the mass of a stable WIMP
within a certain mass range, $m_{\rm min} \leq m_S \leq m_{\rm max}$, 
which we refer to as the Lee-Weinberg window \cite{Lee:1977ua}.
This window is model-dependent and typcally extends from a few GeV to a few TeV. 
If the neutralino is the lightest stable supersymmetric particle, $\mmi \simeq 5$ GeV 
\cite{Belanger:2003wb} but in other models of dark matter $\mmi$ can be 
lowered \cite{BPtV,Fayet:2004bw}.

Recently, WIMPs with masses in the GeV and sub-GeV range have 
been proposed as a solution to certain problems in astrophysics and cosmology. 
For example, sub-GeV WIMPs can produce a high yield of positrons in 
the products of WIMP annihilation near the centres of galaxies 
\cite{Boehm:2003bt}, which may account for 511 KeV photons 
observed recently in the emission from the Galactic bulge 
\cite{Jean:2003ci}. 
GeV-scale WIMPs are also preferred in models of self-interacting
dark matter \cite{Spergel:1999mh} that can rectify the problem with
over-dense galactic centers predicted in numerical 
simulations with non-interacting cold dark matter. 

Dedicated underground experiments have little sensitivity 
to dark matter in the GeV and sub-GeV range. 
Direct detection of the nuclear recoil from the scattering of
such relatively light particles 
is very difficult because of the rather low energy transfer to nuclei, 
$\Delta E \sim v^2 m_S^2/m_{\rm Nucl} \la 0.1$ KeV, 
which significantly weakens experimental bounds on scattering rates 
below $m_S$ of few GeV, especially for heavy nuclei. 
Indirect detection via energetic neutrinos from the annihilation in the 
centre of the Sun/Earth is simply  not possible in this mass range 
because of the absence of directionality. Therefore, the direct 
production of dark matter particles in particle physics experiments stands out 
as the most reliable way of detecting WIMPs in the GeV and sub-GeV mass range. 

The purpose of this work is to prove that $B$ decays can be an effective probe 
of dark matter near the lower edge of the Lee-Weinberg window. $K$ decays can also be 
used for this purpose, but $B$ decays have far greater reach, up to $m_S\sim 2.6$ GeV. 
In particular, we show that pair production of WIMPs in the decays $B\rightarrow K(K^*)SS$ 
can compete with the Standard Model mode 
$B\rightarrow K(K^*)\nu\bar\nu$. In what follows, we analyze in 
detail the ``missing energy'' processes
in the  model of the singlet scalar WIMPs 
\cite{BPtV,Silveira:1985rk,McDonald:1994ex} and use the 
existing data from $B$ physics experiments to put 
important limits on the allowed mass range of scalar WIMPs. 

The main advantage of the singlet scalar model of dark matter is its simplicity,
\bea 
\label{lagr}
-{\cal L}_S= \frac{\lambda_S}{4}S^4+ \frac{m_0^2}{2} S^2+ 
\lambda S^2  H^\dagger H\;\;\;\;\;\;\;\\\nonumber= \frac{\lambda_S}{4}S^4+
\frac12 (m_0^2 + \lambda v_{EW}^2) S^2 + \lambda v_{EW} S^2  h + \frac{\lambda}{2}S^2 h^2,
\eea
where $H$ is the SM Higgs field doublet, $v_{EW} = 246$ GeV is the 
Higgs vacuum expectation value (vev) and $h$ is the field corresponding to the physical 
Higgs, $H = (0,(v_{EW}+h)/\sqrt 2)$. It is important to recognize that the physical 
mass of the scalar $S$ receives contributions from two terms,  
$m_S^2 = m_0^2 + \lambda v_{EW}^2$,
and can be small, even if each term is on the order $\pm O(v_{EW}^2)$. Although admittedly 
fine-tuned, the possibility of low $m_S$ is not {\em a priori} excluded and deserves 
further studies as it also leads to Higgs decays saturated by the invisible channel,
$h\rightarrow SS$ and suppression of all observable modes of Higgs decay at 
hadronic colliders \cite{BPtV}. The minimal scalar model is not a unique 
possibility for light dark matter, which can be introduced more naturally in other models. 
If for example, the dark matter scalar $S$ couples to the $H_d$ Higgs doublet 
in the two-Higgs modification of (\ref{lagr}), $\lambda S^2 H_d^\dagger H_d$, the 
fine-tuning can be relaxed if the ratio of the two electroweak vevs, $\tan\beta = 
\langle H_u\rangle/\langle  H_d \rangle$ is a large parameter. 
A well-motivated case of $\tan \beta \sim 50$ corresponds to $\langle H_d \rangle \sim $ 5 GeV, 
and only a modest degree of cancellation between 
$m_0^2$ and $\lambda \langle H_d\rangle ^2$ would be required to bring $m_S$ in the GeV range. 
More model-building possibilities open up if new particles, other than electroweak gauge bosons 
or Higgses,  mediate the interaction between WIMPs and SM particles. If the mass scale 
of these new particles is smaller than the electroweak scale \cite{Fayet:2004bw},
sub-GeV WIMPs are possible without fine-tuning. 

\begin{figure}\centerline{%
   \psfig{file=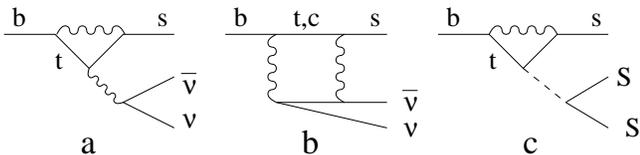,width=8.5cm,angle=0}%
         }
\vspace{0.1in}
\caption{\label{figure:Feyn1} Feynman diagrams which contribute to $B$ meson
decays with missing energy.}
\end{figure}

\section{Pair-production of WIMPs in $B$ decays}

The Higgs mass $m_h$ is heavy compared to $m_S$ of interest, 
which means that in {\em all processes} such as annihilation, pair production, and 
elastic scattering of $S$ particles, $\lambda$ and $m_h$ will enter in  the 
same combination, $\lambda^2 m_h^{-4}$. In what follows, we calculate the 
pair-production of $S$ particles in $B$ decays in terms of two parameters,
$\lambda^2 /m_h^4$ and $m_S$, and relate 
them  using the dark matter abundance calculation, 
thus obtaining the definitive prediction for the signal 
as a function of $m_S$ alone. 

\par
At the quark level the decays of the $B$ meson with missing energy 
correspond to the processes shown in Figure \ref{figure:Feyn1}.
The SM neutrino decay channel is shown in Figure \ref{figure:Feyn1}a and 
\ref{figure:Feyn1}b.
The $b\rightarrow s$ Higgs penguin 
transition,  Figure \ref{figure:Feyn1}c,
produces a pair of scalar WIMPs $S$ in the final state, which likewise
leave a missing energy signal.
In this section, we show that this additional amplitude generates 
$b\rightarrow sSS$ decays that can successfully 
compete with the SM neutrino channel. 

\par A loop-generated $b-s-$Higgs vertex at low momentum transfer
can be easily calculated by differentiating the two-point $b\rightarrow s$
amplitude over $v_{EW}$. We find that to leading order 
the $b \to s h$ transition is given by an effective interaction (see \eg \cite{Willey:1982mc})
\be \label{eq:effint}
{\cal L}_{bsh} = 
\left(\frac{3 g_W^2 m_b m_t^2 V_{ts}^* V_{tb}}{64 \pi^2 M_W^2 v_{EW}} 
\right) \overline{s}_L b_R h +(h.c.).
\label{bsh}
\ee
Using this vertex, Eq. (\ref{lagr}) and 
safely assuming $m_h\gg m_b$, we integrate out the massive Higgs boson
to obtain the effective Lagrangian for the $b\rightarrow s$ transition with 
missing energy in the final state:
\be
{\cal L}_{b\rightarrow s E\!\!\!\!/} = \fr{1}{2}C_{DM} m_b\bar s_L b_R S^2 -
C_\nu \bar s_L \gamma_\mu b_L \bar \nu \gamma_\mu \nu+(h.c.). 
\label{bsE}
\ee
Leading order Wilson coefficients for the transitions with dark matter scalars or 
neutrinos in the final state are given by 
\begin{eqnarray}
\label{CDM}
C_{DM} = \frac{\lambda }{m_h^2} ~\frac{ 3  g_W^2 V_{ts}^* V_{tb}}{32 \pi^2}~x_t
\;\;\;\;\;\;\;\;\;\;\;\;\\
C_\nu = \fr{g_W^2 }{M_W^2} ~\fr{g_W^2V_{ts}^* V_{tb}}{16\pi^2}
\left[\fr{x_t^2+ 2 x_t}{8(x_t-1)}
+\fr{3x_t^2-6x_t}{8(x_t-1)^2}\ln x_t \right],
\nonumber
\end{eqnarray}
where $x_t = m_t^2/M_W^2$. 

We would like to remark at this point that the numerical value of 
$C_{DM}$ is a factor of few larger than $C_\nu$,
\be
\fr{C_{DM}}{C_\nu} \simeq \fr{4.4 \lambda M_W^2}{g_W^2 m_h^2},
\label{ratio}
\ee
if $\lambda m_h^{-2} \sim O(g_W^2 M_W^{-2})$. This happens 
despite the fact that the effective $bsh$ vertex is suppressed relative to 
$bsZ$ vertex by a small Yukawa coupling $\sim m_b/v_{EW}$. 
The $1/v_{EW}$ in (\ref{bsh}) is compensated by a large coupling of $h$ to $S^2$, 
proportional to $\lambda v_{EW}$, and $m_b$ is absorbed into the definition of the 
dimension 6 operator $m_b\bar s_L b_RS^2$.

We concentrate on exclusive decay modes with missing energy, as these
are experimentally more promising than inclusive decays and give
sensitivity to a large range of $m_S$.
A limit on the branching ratio
has recently been reported by \babar\ collaboration, 
${\rm Br}_{B^+\to K^+\nu\bar\nu} < 7.0\times 10^{-5}$ at 90\%
c.l. \cite{Aubert:2003yh}, which improves on a previous
CLEO limit \cite{Browder:2000qr},
but is still far from the SM prediction
${\rm Br}(B\rightarrow K\nu\bar\nu) \simeq (3-5)\times 10^{-6} $
(See, e.g. \cite{Buchalla:2000sk}).
We use the result for ${\cal L}_{b\rightarrow s E\!\!\!\!/}$ 
along with the hadronic form factors determined via 
light-cone sum rule analysis in \cite{abhh} 
and related to the scalar $B\to K$ transition in \cite{Bobeth:2001sq},
to produce the amplitude of $B \to K S S$ decay,
\be
{\cal M}_{B\to KSS} = C_{DM} m_b \frac{M_B^2-M_K^2}{m_b-m_s} f_0(q^2),
\ee
where $q^2 = \hat s =(p_B-p_K)^2$ and the form factor for $B\to K$ transition 
is approximated as $f_0\simeq 0.3 \exp\{0.63 \hat sM_B^{-2} - 0.095 \hat s^2 M_B^{-4}
+0.591 \hat s^3 M_B^{-6}\}$. 

The differential decay width to a $K$ meson and a pair of WIMPs is given by
\be \label{eq:sdw}
\frac{d\Gamma_{B^+\to K^+SS}}{d\hat s} = \frac{x_t^2 C_{DM}^2 f_0(\hat s)^2}{512 \pi^3}  
\frac{I(\hat s,m_S)m_b^2 (M_B^2-M_K^2)^2}{M_B^3(m_b-m_s)^2},
\ee
where $I(\hat s,m_S)$ reflects the available phase space, 
$$
I(\hat s,m_S)= [\hat s^2 - 2\hat s(M_B^2 +M_K^2)+(M_B^2-M_K^2)^2]^{\fr{1}{2}}
[1-4m_S^2/\hat s]^{\fr{1}{2}}.
$$
From Eq. (\ref{eq:sdw}) and the prediction for the SM neutrino channel, 
we obtain the total 
branching ratio for the $B^+$ to $K^+$ decay with missing energy in the final state, 
\begin{eqnarray}
{\rm Br}_{B^+\to K^+ + E\!\!\!\!/}= {\rm Br}_{B^+\to K^+\nu\bar\nu} + 
{\rm Br}_{B^+\to K^+SS} \nonumber 
\\
\simeq 4\times 10^{-6} + 2.8 \times 10^{-4} \kappa^2 F(m_S).
\label{totalbr}
\end{eqnarray}
Eq. (\ref{totalbr}) uses the parametrization of $\lambda^2 m_h^{-4}$,
\be
 \coupling^2 \equiv \lambda^2 \left( \frac{100 ~{\rm GeV}}{m_h} \right)^4,
\ee
and the available phase space as a function of the unknown $m_S$, 
$$
F(m_S)=\int_{\hat s_{min}}^{\hat s_{max}}\!\!\!\!\!\!\! 
f_0(\hat{s})^2 I(\hat s,m_S)~d\hat s ~\left[
\int_{\hat s_{min}}^{\hat s_{max}} \!\!\!\!\!\!\! 
f_0(\hat{s})^2 I(\hat s,0)~d\hat s~\right ]^{-1}
$$
Notice that $F(0) = 1$ and $F(m_S) = 0$ for $ m_S>\fr{1}{2}(m_B-m_K)$ by construction. 
Similar calculations can be used for the decay $B \to K^* SS$, 

\be
{\rm Br}_{B^+\to K^{+*} + E\!\!\!\!/}
\simeq 1.3\times 10^{-5} + 4.3 \times 10^{-4} \kappa^2 F(m_S).
\ee

\noindent
with an analogous form factor.

For light scalars, $m_S\sim$ few 100 MeV, and $\kappa \sim O(1)$ 
the decay rates with emission of dark matter particles are $\sim 50$ 
times larger than the decay with neutrinos in the final state! This is partly due to a 
larger amplitude, Eq. (\ref{ratio}), and partly due to phase space integral that is 
a factor of a few larger for scalars than for neutrinos if $m_S$ is small.  

\section{Abundance calculation and Comparison with Experiment}

\par
 The scalar coupling constant $\lambda$ and the scalar mass $m_S$ 
are constrained by the observed abundance of 
dark matter. For low $m_S$, as shown in \cite{BPtV}, the acceptable value of $\kappa$ 
is  $\kappa\sim O(1)$. 
Here we refine the abundance calculation for 
the range $0< m_S< 2.4$ GeV in order to obtain a more 
accurate quantitative prediction for $\kappa$. 
The main parameter that governs the energy density 
of WIMP particles today, which we take to be equal to the observed value of 
$\Omega_{DM} h^2\sim 0.13$ \cite{Spergel:2003cb},
 is the average of their annihilation cross section at the 
time of freeze-out. This cross section multiplied by the relative 
velocity of the annihilating WIMPs is fixed by $\Omega_{DM}$ and 
can be conveniently expressed \cite{BPtV} as
\begin{eqnarray}
\label{approx}
\sigma_{\rm ann} \; v_{rel} &=& 
\frac{8\vew^{2}\lambda^{2}}{m_h^{4}} \times\left(\lim_{\mht \to 2 \ms} 
\mht^{-1}\Gamma_{\sss \tilde{h} X}\right)
\end{eqnarray}
Here $\Gamma_{\sss \tilde{h} X}$ denotes the partial rate for the decay, 
$\tilde{h} \to X$, for a virtual Higgs, $\tilde{h}$, with the mass of 
$\mht = 2 E_S\simeq 2m_S$. Notice that Eq. (\ref{approx}) contains the same 
combination $\lambda^2 m_h^{-4}$ as (\ref{totalbr}). The zero-temperature width 
$\Gamma_{\sss \tilde{h} X}$ was extensively 
studied two decades ago in conjunction with searches for light Higgs
\cite{voloshin,raby-west,truong-willey}.

\begin{figure}
\psfig{file=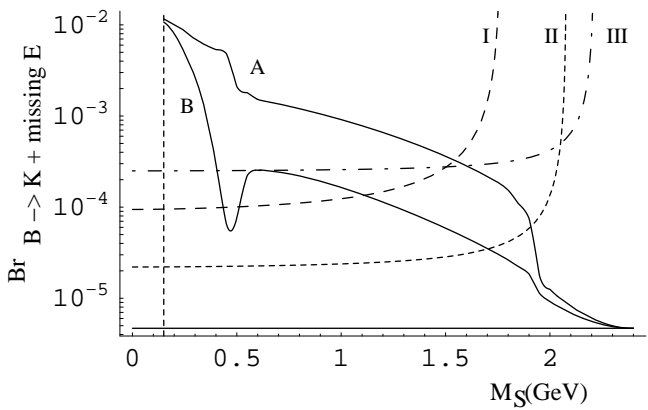,width=8.5cm,angle=0}%
\caption{\label{figure:BranchRatio} Predicted branching ratios for 
the decay $B \to K +$ {\it missing energy}, with current limits 
from \babar\ (I) \cite{Aubert:2003yh}, CLEO (III) \cite{Browder:2000qr} 
and expected results from \babar\ (II). Parameter space above curves I and III is excluded.
The horizontal line shows the SM $B\to K\nu\bar \nu$ signal. Parameter 
space to the left of the vertical dashed line is
also excluded by $K^+ \to \pi^+E\!\!\!\!/ $.}

\end{figure}

For $m_S$ larger than 
$m_\pi$ the annihilation to hadrons dominates the cross section, which
is therefore prone to considerable uncertainties. At a given value of $m_S$, 
we can predict $\Gamma_{\sss \tilde{h} X}$ within a certain range
that reflects these uncertainties. With the use of (\ref{approx}), this prediction
translates into the upper (A) and  the lower (B) bounds on  $\kappa(m_S)$, which we insert 
into Eq. (\ref{totalbr}) and plot the resulting ${\rm Br}_{B^+\to K^+ + E\!\!\!\!/}$
in Figure \ref{figure:BranchRatio}. 

In the interval $150~{\rm MeV}
\le m_S \la 350 $ MeV the annihilation cross section is 
dominated by continuum pions in the final state, and can be calculated 
with the use of low-energy theorems \cite{voloshin} to good accuracy.
Requiring $\kappa^2 < 4\pi$ allows to determine the lower end of the 
Lee-Weinberg window in our model, $\mmi \sim 350$ MeV.
In the interval 
$  350~{\rm MeV} - 650 ~{\rm MeV}$ the strangeness threshold opens up, and 
annihilation into pions via the $s$-channel $f_0$ resonance become important. 
The strength of this resonance and its width and position at 
freeze-out temperatures, $T \sim (0.05-0.1) m_S$, are uncertain. 
Curve B in this domain of Figure \ref{figure:BranchRatio}
assumes the $f_0$ resonance is completely insensitive to
thermal effects and has the {\em minimum} width quoted by PDG 
\cite{Hagiwara:2002fs} which maximizes $\Gamma_{\sss \tilde{h} X}$, 
whereas curve A corresponds to a complete smearing of the $f_0$ resonance 
by thermal effects and much lower value of $\Gamma_{\sss \tilde{h} X}$.
Above $m_S = $1 GeV, curve B takes into account the annihilation into 
hadrons mediated by $\alpha_s (G_{\mu\nu})^2$ {\em with} the two-fold 
enhancement suggested by charmonium decays \cite{voloshin} whereas curve A
uses the perturbative formula. The charm threshold is treated simply 
by the inclusion of open charm quark production at 
a low threshold ($m_c \simeq 1.2$ GeV) 
in B curve and at a high threshold ($m_S>m_D$) in curve A. 
Both curves include the $\tau$ threshold. 
There are no tractable ways of calculating the cross section 
in the intermediate region $650 ~{\rm MeV} \la m_S \la 1$ GeV. 
However, there are no particular reasons to believe that the 
annihilation into hadrons will be significantly enhanced or 
suppressed relative to the levels in adjacent domains.
In this region, we interpolate between high- and low-energy sections of curves A and B. 
Thus, the parameter space consistent with the required cosmological abundance of
$S$ scalars calculated with generous assumptions 
about strong interaction uncertainties 
is given by the area between the two curves, A and B. 

Figure 2 presents 
the predicted range of ${\rm Br}_{B^+\to K^+ + E\!\!\!\!/}$ as a 
function of $m_S$ and is the main result of our paper. 
The SM ``background'' from $B \to K\nu\bar \nu$ decay is subdominant
everywhere except for the highest kinematically allowed domain of $m_S$.
To compare with experimental results \cite{Aubert:2003yh,Browder:2000qr}, 
we must convert 
the limit on ${\rm Br}_{B^+\to K^+\nu\bar\nu}$ to a more appropriate bound on 
${\rm Br}_{B^+\to K^+ + E\!\!\!\!/}$ according to the following procedure. 
We multiply the experimental limit of $7.0\times 10^{-5}$ by a ratio
of two phase space integrals, $F(m_S,\hat s_{min})/F(m_S,\hat s_{exp})$,
where $s_{exp}$ is determined by the minimum Kaon momentum considered
in the experimental search, namely $1.5$ GeV. 
This produces an exclusion curve, 
nearly parallel to the $m_S$ axis at low $m_S$, and almost vertical near the 
experimental kinematic cutoff. 
The current \babar\ results (curve I) exclude 
$m_S< \bbounda$ MeV, as well as the region $\bboundb$ MeV$< m_S <$ $\bboundc$ GeV, 
and probe the allowed parameter space 
for dark matter up to $m_S \sim \bboundd$ GeV. 
Generalized model with $N$ component dark matter scalar
gives $N^2$-fold increase in the branching ratio \cite{BPtV}, and thus 
greater sensitivity to $m_S$.

The $B$ factories will soon have larger data samples and can extend the
search to lower Kaon momenta.  The level of sensitivity expected 
from an integrated luminosity ${\cal L}$ of 250 fb$^{-1}$ and momentum cutoff 
of $1$ GeV is shown by curve II, which assumes that the sensitivity
scales as ${\cal L}^{-1/2}$, as suggested by the analysis
in \cite{Aubert:2003yh}.  In reality, the experimental limit
will extend to Kaon momenta below 1~GeV where the sensitivity will
gradually degrade due to increasing backgrounds; however, we expect
the implication of curve II to remain valid, namely 
that the $B$ factories will probe scalar dark matter up to 2~GeV.

If $m_S\la 150$ MeV, the decay $K^+ \to \pi^+ SS$ becomes
possible. The width for this decay can be easily calculated in a similar 
fashion to $b\to s$ transition. The concordance of the observed number of events 
with the SM prediction \cite{Adler:2001xv} 
rules out scalars in our model with $m_S< 150$ MeV. This 
exclusion limit is shown by a vertical line in Figure 3. 
It is below $\mmi$ of 350 MeV. 


To conclude, we have demonstrated that the 
$b\to s$ transitions with missing energy in the final state can be
an efficient probe of dark matter when pair production 
of WIMPs in $B$ meson decays is kinematically allowed. 
In particular, we have shown that the minimal
scalar model of dark matter with the interaction mediated by the Higgs 
particle predicts observable rates for $B^{+}\to K^{+}$ and missing energy.
A large portion of the 
parameter space with $m_S\la 1$ GeV is already 
excluded by current \babar\ limits. New experimental data should
probe a wider range of masses, up to $m_S \sim 2$ GeV. The limits
obtained in this paper have important implications for Higgs searches, 
as the existence of relatively light scalar WIMPs leads to the Higgs 
decays saturated by invisible channel. Given the astrophysical motivations 
for GeV and sub-GeV WIMPs combined with insensitivity of dedicated dark matter searches
in this mass range, it is important to extend the analysis of $b\to s$ transition with 
missing energy onto other models of light dark matter. 

We thank Misha Voloshin for valuable discussion. 
This research is supported in part by NSERC of Canada and PPARC UK.

\bibliography{bjkp_rev}

\end{document}